\documentclass[11pt]{aastex}

\begin{document}

\bibliographystyle{apj}

\shorttitle{Millimeter Emission in GQ Lup}

\slugcomment{revised for AJ: November 18, 2009}

\title{Millimeter Dust Emission in the GQ Lup System}

\author{Yu Dai\altaffilmark{1,2}, 
David J. Wilner\altaffilmark{2}, 
Sean M. Andrews\altaffilmark{2,3},
and Nagayoshi Ohashi\altaffilmark{4}
}
\altaffiltext{1}{Physics Department, Boston College, 140 Commonwealth Avenue, 
Chestnut Hill, MA 02467}
\altaffiltext{2}{Harvard-Smithsonian Center for Astrophysics, 60 Garden Street,
Cambridge, MA 02138}
\altaffiltext{3}{Hubble Fellow}
\altaffiltext{4}{Institute of Astronomy and Astrophysics, Academia Sinica, 
P.O. Box 23-141, Taipei 106, Taiwan}
\email{ydai$@$cfa.harvard.edu}

\begin{abstract}
We present Submillimeter Array observations of the GQ Lup system at 
1.3~millimeters wavelength with $0\farcs4$ ($\sim$60~AU) resolution.
Emission is detected from the position of the primary star, GQ Lup~A, 
and is marginally resolved. No emission is detected from the substellar 
companion, GQ Lup~B, $0\farcs7$ away.
These data, together with models of the spectral energy distribution, 
suggest a compact disk around GQ Lup~A with mass $\sim 3$~M$_{Jup}$,
perhaps truncated by tidal forces. 
There is no evidence for a gap or hole in the disk that might be the 
signature of an additional inner companion body capable of scattering 
GQ Lup~B out to $\sim100$~AU separation from GQ Lup~A.
For GQ Lup~B to have formed {\it in situ}, the disk must have 
been much more massive and extended. 
\end{abstract}


\section{Introduction}

The classical T Tauri star GQ Lup~A (spectral type K7, age $\sim 1$~Myr) 
has received considerable attention since the detection by direct imaging 
in the near-infrared of a companion, GQ Lup~B, at projected separation 
$\sim0\farcs7$ ($\sim100$~AU at 150 pc) with a mass originally claimed to be 
perhaps as low as the planet Jupiter \citep{neu05}.
Subsequent spectroscopy and analysis suggested a higher mass for the 
companion, most likely a brown dwarf in the range of 10 to 40 M$_{Jup}$ 
\citep[e.g.][]{mug05,gue06,mce07,sei07,mar08,neu08}, making it one of a 
small but growing class of pre-main-sequence stars with substellar companions 
at $\sim100$~AU separations \citep{luh06,sch08}.

Such wide-separation, high-mass ratio companions are difficult to form 
{\it in situ} within circumstellar disks either by the standard 
core-accretion process or by the gravitational instability mechanism
\citep{bos06,deb06,ver09}.
Observations of the disk(s) in these young systems may offer clues to their 
formation. While the GQ Lup system exhibits all of the usual pre-main-sequence 
disk signatures, including optical emission lines due to gas accretion and 
thermal dust emission from infrared to millimeter wavelengths \citep{nue97}, 
it is not known whether the bulk of the disk mass is associated with the 
primary star, the substellar companion, or resides in a circumbinary structure.
If GQ Lup~B formed as a ``planet'' within a circumstellar disk, 
for example, then one might expect disk material to be present at large 
radii, beyond its orbit, ultimately evolving into a debris disk like those 
surrounding the recently imaged bodies 
orbiting HR~8799 \citep{mar08} and Fomalhaut \citep{kal08}.

Interferometric imaging of thermal dust continuum emission at millimeter 
wavelengths offers a way to trace the location of cool material in the 
GQ Lup system. In this paper, we present 1.3 millimeter observations 
with sufficient angular resolution to separate the primary and the companion,
which reveal a compact disk surrounding GQ Lup~A. We model the disk emission
to constrain physical properties and GQ Lup~B formation scenarios.

\section{Observations}
\label{sec:obs}
We observed GQ Lup at 1.3~millimeters wavelength with 
Submillimeter Array (SMA) 
\footnote{The Submillimeter Array is a joint project between the
Smithsonian Astrophysical Observatory and the Academica Sinica Institute of
Astronomy and Astrophysics and is funded by the Smithsonian Institution and 
the Academica Sinica.} on 
2007 June 19 as a short filler project, at the start of a set of antenna
moves from the very extended configuration to a compact configuration.
The six available antennas provided projected baselines from 30 to 430 meters.
The two hour observation was performed in excellent weather conditions,
with 225 GHz atmospheric opacity of 0.05, as measured at the nearby 
Caltech Submillimeter Observatory.
The correlator provided 2~GHz bandwidth in two sidebands, with a central 
LO frequency of 225~GHz. Observations of the two quasars J1517-243 and 
J1454-377 were interleaved with GQ Lup
(R.A. = $15^h49^m12\fs138$, DEC =$-35\degr39\arcmin03\farcs9$, J2000).  
Passband calibration was
accomplished with observations of the strong source 3c273.
Complex gain calibration was done using J1517-243. The position derived for
J1454-377, within $0\farcs2$ of its catalog position, provides an empirical 
measure of the astrometric accuracy that should also hold for GQ Lup. 
The absolute flux scale was set with reference to the standard calibrator 
MWC349, assumed to be 1.65 Jy at 225 GHz, accurate to better than 20\%.
All of the calibration was performed with the IDL based MIR software.
Subsequent imaging was performed using standard routines in the MIRIAD package.

\section{Results and Analysis}
\label{sec:result}
\subsection{1.3~mm Continuum}
\label{sec:continuum}
\begin{figure}[h]
\includegraphics[scale=0.95,angle=0]{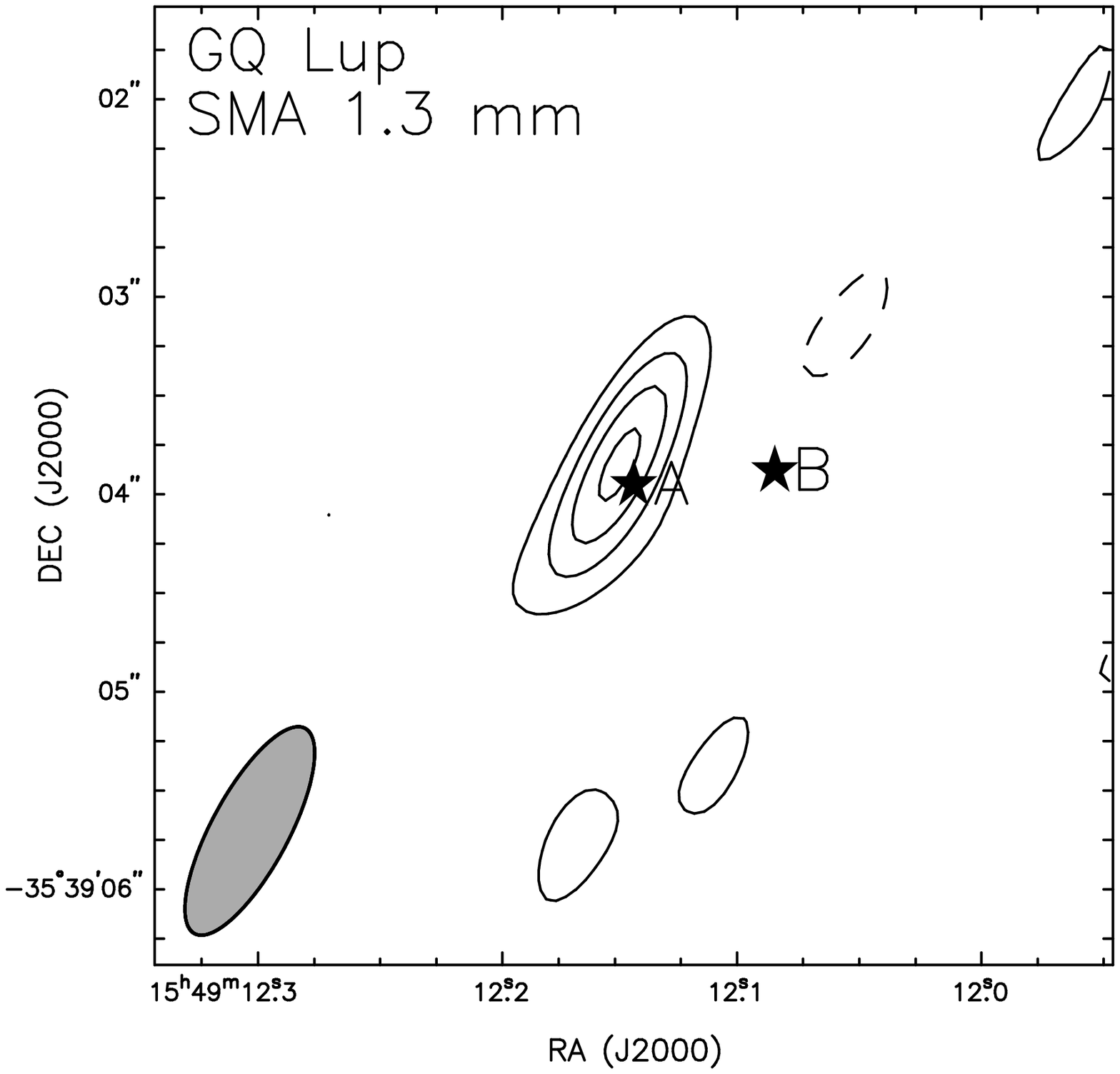}
\figcaption{
SMA image of the 1.3 millimeter continuum emission in the GQ Lup system.
The contour levels are $-2,2,4,6,...\times2.8$~mJy (the rms noise level).
Negative contours are dashed. The star symbols mark the positions of 
the star GQ Lup~A and substellar companion GQ Lup~B.
The millimeter emission peak is associated with GQ Lup~A.
The filled ellipse indicates the
$1\farcs2\times0\farcs4$, P.A. $-28^{\degr}$ synthesized beam.
\label{fig:image}}
\end{figure}

Figure~\ref{fig:image} shows the image of the GQ Lup 1.3 millimeter
continuum emission, with synthesized beam $1\farcs2 \times 0\farcs4$ at 
position angle $-28\degr$ using data from both sidebands and natural
weighting to obtain the best signal-to-noise ratio.
The beam is highly elliptical due to the low declination of the source 
combined with the limited $(u,v)$ coverage provided by the short span of 
the observations. Nonetheless, examination of the image 
shows that the emission clearly peaks at the primary star, GQ Lup~A,
within the astrometric uncertainty, and 
no significant emission is detected from the substellar companion, GQ Lup~B. 
(Note that any unaccounted-for orbital motion of GQ Lup~B is very small, 
well below the resolution of these observations \citep{mug05}.)
The flux density determined either by integrating in a box around the 
emission in the image, or by fitting the visibilities with a Gaussian, 
is $25\pm3$~mJy (statistical error only).
We identify the emission with a compact disk surrounding GQ Lup~A.

The SMA flux density is consistent, within the uncertainties, with the 
SEST single dish bolometer measurement of 38$\pm$7 mJy \citep{nue97}.
The agreement is likely better than suggested by a strict comparison of 
the values, considering that the effective frequency of the $\sim50$~GHz 
wide SEST bandwidth is likely 240 to 250 GHz on account of the steeply 
rising dust spectrum \citep[e.g.][]{car01}, which biases the measurement
to a higher value (at least 20\% for S$_\nu \propto \nu^3$).
Given this consistency, we conclude that no significant
emission has been missed due to spatial filtering by the interferometer.

The left panel of Figure~\ref{fig:SED} shows the visibility amplitude for 
the SMA data
as a function of baseline length, averaged in concentric circular annuli, 
appropriate for a disk viewed face-on. 
The disk inclination is unknown and cannot be determined from these data,
but if it is similar to the inclination of the binary system of 
$27\pm5^{\degr}$ \citep{bro07,hug09}, then the face-on approximation is 
reasonable. 
While the bins are coarse and the signal-to-noise per bin is modest
($4-5\sigma$), the slight trend of decreasing amplitude at longer
baselines suggests that the emission from GQ Lup is unresolved or 
marginally resolved.
To estimate the atmospheric seeing effect on GQ Lup, we fit the 
calibrated visibilities of J1454-377, which has a similar angular 
separation from J1517-243 as GQ Lup (albeit in a different direction), 
and obtain $\lesssim0\farcs10$ fwhm.  The seeing is a minor 
contributor to the visibility amplitude trend. 
Note that if we make the extreme assumption that the source emission is 
optically thick, then the observed flux density implies a minimum size.
For a face-on disk with characteristic temperature 30 K, the minimum 
radius is $\sim10$~AU (diameter $\sim0\farcs13$).

\subsection{SED}
\label{sec:SED}
The right panel of Figure~\ref{fig:SED} shows the spectral energy distribution 
(SED) of the 
GQ Lup system, including the new SMA datum and values from the literature.  
The optical {\em B,V,R,I} data are from \citet{her94}, dereddened assuming 
A$_V$=0.5 and the \citet{mat90} extinction law.  
The near infrared {\em J,H,K} data are from the
{\em Two Micron All Sky Survey} \citep{skr06}.
At longer infrared wavelengths, the spectrum from 5 to 40 $\mu$m is
taken from the public archive of the 
{\em Spitzer Space Telescope} `c2d' legacy program \citep{eva03,kes06}, 
and data at 12, 25, and 60 $\mu$m are from the 
{\em IRAS} Faint Source Catalog \citep{mos90}.
The {\em IRAS} data appear to be systematically higher than the 
{\em Spitzer} data, more than what is readily explained by the presence 
of silicate features in the spectrum near 10 and 20 $\mu$m or by the 
nominal instrumental calibration uncertainties.

\subsection{Disk Models}
\label{sec:models}

To constrain the disk properties, we have calculated a series of models 
using the radiative transfer code 
{\tt RADMC} \citep{dul04} employing an approach similar to \citet{and09}. 
Any constraints on the disk parameters must be considered in the context 
of the model assumptions, given the limitations of the data available.

A natural physical model for the GQ Lup system includes a circumprimary
disk with a sharp outer edge, as might result from tidal truncation by
the secondary. Alternative models are possible but are not considered.
In detail, we assume that the (dust) disk density structure is parameterized 
by a radial surface density power-law truncated at an outer radius, $R_{d}$, 
i.e.  $\Sigma = \Sigma_0 (R/R_0)^{-p}, R_{in} < R < R_{d}$,
and a vertical scale height power-law $H = H_0 (R/R_0)^{1+\psi}$
with a ``puffed-up'' inner rim \citep[as in][]{dul04}.
The disk temperature structure is then calculated self-consistently, 
assuming stellar irradiation is the only heating source. 
The accretion rate of $3\times10^{-9}$~M$_{\odot}$ derived by \citet{hug09}
provides negligible heating compared to the star.
For these models, we use the stellar spectral type and dereddened 
optical data to fix the stellar effective temperature, $T_* = 4060$~K 
and luminosity, $L_* = 1.5$~L$_{\odot}$, and fix $R_{in}$ at the dust 
sublimation radius of 0.09~AU (for a sublimation temperature of 1500~K).
We also fix the surface density power law index, $p=1$. 
This power law index is compatible with constant $\alpha$ irradiated 
accretion disks away from the disk boundaries \citep{dal98}, as well as 
resolved millimeter observations of T Tauri star disks \citep{wil00,and07}.
We adopt the dust opacities used by \citet{and09} and a standard gas-to-dust 
ratio of 100. The resulting mass opacity for (dust + gas) at 1.3~millimeters 
of $0.0226$~cm$^2$g$^{-1}$ is very similar to the commonly adopted value of 
$0.02$~cm$^2$g$^{-1}$ of \citet{bec90}.
To make comparisons with the resolved millimeter data, we use a 
Gaussian with $0\farcs1$ fwhm
to approximate the atmospheric seeing effect,
as estimated empirically from the calibrator observations.

\begin{figure}[h]
\includegraphics[scale=1.0,angle=0]{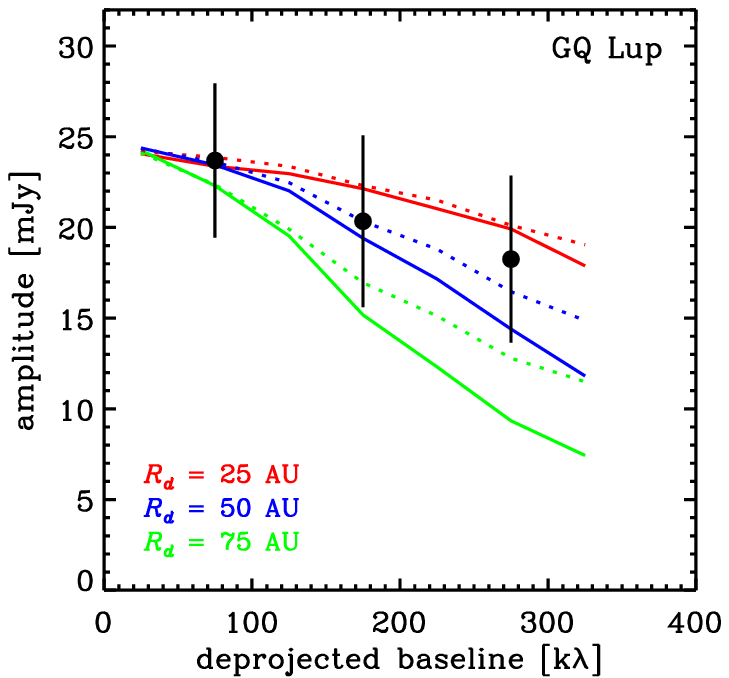}
\includegraphics[scale=1.0,angle=0]{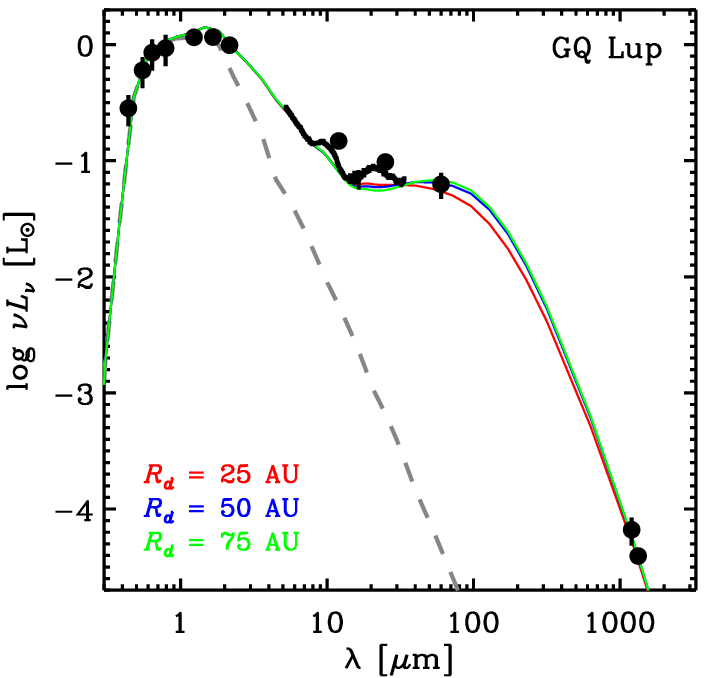}
\figcaption{
{\em (left)} The 1.3 millimeter visibility amplitude as a function of  
baseline length, averaged in three concentric circular annuli.
The expected amplitude for zero signal is $\sim5$~mJy in each bin.
The three solid curves are from the disk model fits to the SED 
described in \S\ref{sec:models}, assuming p=1 and disk radius values 
$R_d = 25, 50, 75$~AU (in red, blue, and green, respectively).  
The three dashed curves are from the corresponding models for p=1.5.
{\em (right)} The spectral energy distribution of GQ Lup, including the 
1.3~millimeter SMA measurement and values from the literature.
The stellar photosphere is indicated by the dashed line. The three solid 
curves are from the disk model fits, as in the left panel.
\label{fig:SED}}
\end{figure}

The three solid curves in the right panel of Figure~\ref{fig:SED} show 
the best-fit models to 
the SED for three values of the disk radius, $R_d = 25, 50, 75$~AU. 
Note that the silicate features near 10 and 20 $\mu$m were excluded 
from the fit, as were the discrepant {\em IRAS} 12 and 25~$\mu$m data.
The fitted values for the parameters that describe the disk vertical 
structure are nearly identical in the three models, as these parameters are 
constrained essentially entirely by the shape of the infrared SED. 
In each of the three models, the flaring angle is characterized by 
$\psi=0.2$, the scale height at 25~AU is 2.35~AU, and an inner rim 
scale height enhanced by 40\% is needed to account for the detailed 
shape of the mid-infrared spectrum relative to the near-infrared emission.
The variations of the disk radius over the range explored have a 
negligible effect on the SED, as expected \citep{ada88}. However, the 
corresponding model curves in the left panel of Figure~\ref{fig:SED} 
show that the disk radius 
has a dramatic effect on the 1.3 millimeter visibilities, in particular 
at the longer baselines that probe the relevant size scales.
The disk models with $R_d = 25$ and $R_d = 50$~AU bracket the data points, 
while the model with $R_d = 75$~AU produces insufficient emission at 
baselines $\gtrsim200$~k$\lambda$. 

The constraint on $R_d$ is sensitive to the model assumptions. In particular, 
there is a well-known degeneracy between the prescribed radial fall-off 
in the disk surface density distribution and the outer radius \citep{mun96}. 
To illustrate an alternative, we have calculated best-fit models that assume 
p=1.5, a steeper surface density distribution, for $R_d = 25, 50, 75$~AU. 
The vertical structure parameters and SED fits in these models are nearly 
identical to the models that assume p=1. The main effect of the steeper 
radial surface density distribution is a less steep decline in the visibility 
function, for fixed outer radius, as shown by the dashed curves in the 
left panel of Figure~\ref{fig:SED}. As expected, a slightly larger 
value of $R_d$ is favored compared to the p=1 models.  A much more extreme 
fall-off of the surface density would be required to obtain $R_d$ as large
as 100~AU or larger. Given the quality and noise of the millimeter data, 
a more complete exploration of disk model parameter space is not warranted.

The disk masses in the three p=1 models in Figure~\ref{fig:SED} are 
0.0025, 0.0029, and 0.0032 M$_{\odot}$, respectively.  
These values are similar to the value of 0.0043~M$_{\odot}$ crudely 
estimated by \citet{nue97} from the 1.3~millimeter flux and an average 
dust temperature of 30~K, assuming the \citet{bec90} mass opacity.
The uncertainty in the disk mass estimates is dominated by the adopted mass
opacity, which depends on the dust properties, including grain sizes, shapes
and composition, as well as the assumed interstellar gas-to-dust ratio.
The systematic uncertainties are significant. Plausible models for dust
properties in disks, for example including the effect of grain growth beyond 
millimeter sizes, give rise to perhaps an order-of-magnitude range in the 
dust opacity at 1.3~millimeters \citep[see e.g.][]{dal01,dra06}.

\section{Discussion}
\label{sec:discussion}
A comprehensive set of scenarios for the origin of widely separated,
low mass companions like GQ Lup~B has been considered by \citet{deb06}, 
including (a) formation as planet {\it in situ} at $\sim100$~AU either 
by core accretion or by gravitational instability, (b) displacement of 
GQ Lup~B to a wider orbit from a formation site much closer in to the 
primary, through stellar encounter, migration, or planet-planet scattering, 
and (c) formation as a brown dwarf by cloud fragmentation or capture. 
While there are arguments for and against each of these formation 
scenarios, and the true mass and nature of GQ Lup~B remain in dispute, 
the observations of disk emission provide some new information. 

For {\it in situ} formation of GQ Lup~B as a planet to be viable, the 
disk surrounding GQ Lup~A must have extended to larger radii in the past, 
with sufficient mass at $\sim100$~AU. 
The detected disk is compact and low mass, only $\sim$3 M$_{Jup}$ in the 
model with $R_d = 50$~AU described in \S\ref{sec:SED}. 
We can extrapolate the disk models to estimate the mass reservoir that 
might have been available for planet formation in the outer disk. 
For a surface density deceasing with radius as $R^{-1}$, a disk extending 
to 250~AU, $5\times$ larger, would have a mass $5\times$ higher. 
In this example, the available disk mass of 12~M$_{Jup}$ beyond 50~AU 
would be only marginally sufficient to explain GQ Lup~B, even 
if all of this disk mass were incorporated into the companion. 
There is no evidence for circumstellar material beyond the orbit of 
GQ Lup~B, and it seems unlikely that the outer disk mass could have 
been substantially larger than this estimate (unless the adopted
millimeter mass opacity were to underestimate the disk mass by a
substantial amount).  Circumstellar disks with 
much larger outer radii do exist, e.g. IM Lup with radius $\sim900$~AU 
in the same cloud complex \citep{van07,pan09}, but these are unusual.
Of course, the circumstellar disk may have been larger and/or more massive 
(and probably gravitationally unstable) at a much earlier evolutionary stage 
when the central protostar was forming. In that environment, the issue 
would be how continued infall and accretion could be quenched to allow the 
companion to remain at substellar mass.

Mechanisms that require moving GQ Lup~B outward to $\sim100$~AU from 
closer in would have an effect on the inner disk, which emits primarily 
in the mid-infrared.
The SED shows no significant flux deficit relative to a continuous disk
that would be indicative of a gap or hole produced by another massive 
companion orbiting at $<10$~AU radius that could have interacted with 
GQ Lup~B to drive an outward migration or to scatter it outward.  
Moreover, such a dynamical interaction 
likely would leave the hypothetical inner planet on an eccentric orbit, 
disrupting the inner disk. The SED also provides no evidence for any major 
disturbance in the disk resulting from a (rare) close stellar encounter 
capable of radically changing the orbital parameters of GQ Lup~B.

The millimeter emission from the GQ Lup system is consistent with the 
flux densities found in surveys of classical T~Tauri binary systems 
with separations in the range of $\sim1 - 100$~AU \citep{jen96,and05}, 
with a relatively low value readily explained by tidal truncation of 
individual circumstellar disks and a paucity of massive circumbinary disks.
Representative calculations of disks in binary systems by \citet{art94} 
show truncation of the circumprimary disk at $\lesssim0.5a$, where $a$ 
is the orbital semi-major axis, together with truncation of the 
circumsecondary disk more severe by a factor of a few. 
While the GQ Lup mass ratio is more extreme than considered by \cite{art94}
and the outer disk radius derived from the resolved millimeter observations 
is model dependent, the data are qualitatively consistent with expectations 
for tidal truncation. The SED and millimeter data show directly the presence  
of a circumprimary disk, and the millimeter flux limit together with the 
suggestion of H$\alpha$ emission \citep{mar07} could indicate a significantly 
smaller circumsecondary disk.
The observations presented here limit the disk mass around GQ Lup~B to be 
approximately $3\times$ less than around GQ Lup~A, or $<1$~M$_{Jup}$.

More sensitive dust continuum observations could show whether or not 
the disk around GQ Lup~A shares a common inclination and orientation with 
any putative disk around GQ Lup~B, or with the GQ Lup~A-B orbit.  
As more substellar companions to pre-main-sequence stars are identified 
by direct imaging, deep millimeter observations will reveal more about 
their formation processes and their enigmatic relationship to planets.

\acknowledgments{
We thank A. Meredith Hughes for insightful comments, Charlie Qi for 
assistance with SMA data reduction, and Kees Dullemond for providing the 
radiative transfer code.  
We thank the SMA staff for scheduling the brief filler observation 
that provided the basis for this paper. 
}

{\it Facility:} \facility{Submillimeter Array}

\end{document}